\documentclass[amsmath,amssymb,twocolumn]{revtex4}
\usepackage[parfill]{parskip}    
\usepackage{graphicx}
\usepackage{amsmath}
\usepackage{amssymb}
\usepackage{epstopdf}
\usepackage{color}
\usepackage{graphicx}
\usepackage{pdfpages}

\begin{document}
\title{\bf All static spherically symmetric anisotropic solutions for general relativistic polytropes}
\author{G. Abell\'an$^{1}$, P. Bargue\~no$^2$, E. Contreras$^3$, E. Fuenmayor$^1$}
\affiliation{
$^1$ Centro de F\'{\i}sica Te\'orica y Computacional, Facultad de Ciencias, Universidad Central de Venezuela, AP 20513, 
Caracas 1020-A, Venezuela.
\\
$^2$ Departamento de F\'isica Aplicada, Universidad de Alicante, Campus de San Vicente del Raspeig, 03690 Alicante, Spain.
\\	
$^3$ Departamento de F\'isica. Colegio de Ciencias e Ingenier\'ia. Universidad San Francisco de Quito, 170901 Quito, Ecuador.
\\
}
\begin{abstract}
An algorithm presented by K. Lake to obtain all static spherically symmetric perfect fluid solutions was recently extended by L. Herrera to the interesting case of locally anisotropic fluids (principal stresses unequal). In this work we develop an algorithm to construct all static spherically symmetric anisotropic solutions for general relativistic polytropes. Again the formalism requires the knowledge of only one function (instead of two) to generate all possible solutions. To illustrate the method some known cases are recovered. 
\end{abstract}

\maketitle

\section{Introduction}
The general formalism to study polytropes for anisotropic matter has been presented in recent papers \cite{Herrera2013,Herrera2013a,herrera2014}. The motivations to undertake such a task lie, on the one hand, in the simplicity of the equation of state and the ensuing structure of the Lane-Emden equation, and on the other hand, in the fact that the polytropic equation of state may be applied to a wide range of very different astrophysical scenarios (see Refs \cite{4,5,6,7,8,9,10,11,12,13,14,15,16,17,18,19,20, pr6,pr9,pr10,pr7,pr8,pr3,pr4} and references therein). Also, we know nowadays that the isotropic pressure condition may be too stringent, and furthermore the presence of local anisotropy is caused by a large variety of physical phenomena expected to be present in compact objects \cite{22,23,24,25,26,27,28,29,30,31,32,32a,32b,32c}. Among all possible sources of anisotropy let us mention two of them which might be particularly related to our primary interest: 
(i) the intense magnetic field observed in compact objects such as white dwarfs, neutron stars, or 
magnetized strange quark stars \cite{33,34,35,36,37,37bis} (in some way, the magnetic field 
can be addressed as a fluid anisotropy) and, (ii): the viscosity, which is another source of anisotropy expected 
to be present in neutron stars and, in general, in highly dense matter (see Refs. \cite{43,44,45,46,47,48,49,50}). Also, as it has been recently proved, the isotropic pressure condition becomes unstable by the presence of dissipation, energy density inhomogeneities and shear \cite{LHP}, which can explain the renewed interest in the study of fluids not satisfying the isotropic pressure condition, and justify our present treatment. It is important to note that, although the degree of anisotropy may be small, the effects produced on compact stellar objects may be appreciable. For example, the occurrence of an 
interesting phenomena such as cracking \cite{pr10} may happen even for slight deviations from isotropy.

The theory of polytropes is based on the polytropic equation of state, which in the Newtonian case reads
\begin{eqnarray}\label{ep0}
P\,=\,K \rho_{0}^{\gamma}\,=\,K\rho_{0}^{1+\frac{1}{n}},
\end{eqnarray}
where $P$ and $\rho_{0}$ denote the isotropic pressure and the mass (baryonic) density, respectively. The constants $K$, $\gamma$, and $n$ are usually called the polytropic constant, polytropic exponent, and polytropic index, respectively. When the polytropic constant $K$ is fixed and can be calculated from natural constants, the polytropic equation of state may be used to model a completely degenerate Fermi gas in the nonrelativistic ($n=5/3$) and relativistic limit ($n=4/3$). In this case, polytropes provide a way of modelling compact objects such as white dwarfs and allow to obtain in a rather direct way the Chandrasekhar mass limit \cite{Herrera2013,abellan20}. Now, if in the problem under consideration the degree of compactness of the stellar objects involved is greater than that corresponding to a white dwarf, for which we can still use Newtonian gravity, we have to resort to the general relativistic regime (e.g. neutron stars, strange-quark stars, super–Chandrasekhar white dwarfs). In fact, general relativistic polytropes have been extensively studied (see for example \cite{Herrera2013a,herrera2014,12,13,14,15,16,17,18,19,20,pr6,pr9,pr10,pr7,pr8,pr3,pr4}) and a remarkable well comprehensive framework to describe them for anisotropic fluids was presented in \cite{Herrera2013a}. In the general relativistic anisotropic case, two possible extensions of the above equation of state (\ref{ep0}) are possible, namely
\begin{eqnarray}
&&P_{r}\,=\,K \rho_{0}^{\gamma}\,=\,K\rho_{0}^{1+\frac{1}{n}}\label{ep1}\\
&&P_{r}\,=\,K \rho^{\gamma}\,=\,K\rho^{1+\frac{1}{n}}\label{ep2},
\end{eqnarray}
where $P_{r}$ and $\rho$ denote the radial pressure and the energy density, respectively, both leading to the same equation (\ref{ep0}) in the Newtonian limit. As it happens in the Newtonian anisotropic case, the fact that the principal stresses are unequal produces an underdetermination of the problem, requiring to impose an additional condition. This means that the assumption of either (\ref{ep1}) or (\ref{ep2}) is not enough to completely integrate the field equations, since the appearance of two principal stresses (instead of one) leads to a system of three independent (field) equations and five unknown functions. Thus, in order to integrate the obtained system of equations, we need to provide further information about the anisotropy, inherent to the problem under consideration. 
Also, we can proceed in a different way in order to integrate the set of equations by imposing certain conditions on the metric variables which are of physical relevance.
An example of this procedure is imposing the vanishing of the Weyl tensor, usually referred to as the
conformally flat condition \cite{Herrera2001}. This condition has its own interest, since it has been seen that highly compact configurations may be obtained with the specific distribution of anisotropy created by such a condition \cite{herrera2014}. Other approaches which come from string theory, the Randall-Sundrum model \cite{RS} or $5$-dimensional warped geometries have retaken the interest for other type of conditions, relating radial derivatives of the metric functions, that produce stellar models embedded in a $5$-dimensional flat space-time (embedding class one). These models satisfy the Karmarkar condition \cite{karmarkar} (for recent developments see, \cite{Ktello,tello2,tello3,tello4,Nunez,tello5,tello6}, for example) and its possible to choose one of these metric functions as the one which generates the total solution. In the isotropic case, only two solutions exist satisfying the Karmarkar-condition. However, when anisotropic interior solutions are considered, a large amount of realistic solutions have been found.

Some years ago, Lake \cite{Lake2003} developed an algorithm which produced all static spherically symmetric and 
isotropic perfect fluid solutions based on a single generating function, thus constructing an infinite number of previously
unknown solutions physically relevant. This work was subsequently extended by Lake himself \cite{Lake2004} and by 
Herrera {\it et al.} \cite{Herrera2008} to the
case of locally anisotropic fluids, which describes very reasonably the matter distribution of a great varity of situations
of interest, as commented before (see, for example, Ref. \cite{22} and references therein).  
In this case, the protocol to obtain all the static and anisotropic solutions  of Einstein's equations can be
summarized as follows. Given a line element parametrized as
\begin{eqnarray}\label{linelement}
ds^{2}=e^{\nu(r)}dt^{2}-e^{\lambda(r)}dr^{2}-r^{2}d\Omega^{2}\,,
\end{eqnarray}

from where, considering an anisotropic fluid, the Einstein field equations read
\begin{eqnarray}
-8\pi\rho&=&-\frac{1}{r^{2}}+e^{-\lambda}\left(\frac{1}{r^{2}}-\frac{\lambda'}{r}\right),\label{eq1}\\
8\pi P_{r}&=&-\frac{1}{r^{2}}+e^{-\lambda}\left(
\frac{1}{r^{2}}+\frac{\nu'}{r}\right),\label{eq11}\\
8\pi P_{\perp}&=& \frac{e^{-\lambda}}{4}
\left(2\nu'' +\nu'^{2}-\lambda'\nu'+2\frac{\nu'-\lambda'}{r}
\right)\label{eq111},
\end{eqnarray}
we introduce the variables
\begin{eqnarray}
e^{\nu}&=&\exp\!\left[{\int(2z-\frac{2}{r})dr}\right] ,\label{def1}\\
e^{-\lambda}&=&y\,, \label{def2}
\end{eqnarray}
where $z=z(r)$ and $y=y(r)$ are the so--called generating functions. In terms of these functions, we obtain
\begin{equation}
e^{\lambda}\,=\,\frac{z^{2}\exp\!\left[{\int(\frac{4}{r^{2}z}}+2z)dr\right]}
{r^{6}\left[-2\int 
\frac{z(1+\Pi r^{2})\exp\left[{\int(\frac{4}{r^{2}z}}+2z)dr\right]}{r^{8}}dr+C\right]}\;,
\end{equation}

where $C$ is a constant of integration and $\Pi=8\pi(P_{r}-P_{\perp})$. Then, replacing (\ref{def1}) in (\ref{eq1}), (\ref{eq11})
and (\ref{eq111}), we arrive to
\begin{align}
4\pi \rho&=\frac{m'}{r^{2}}\label{rhoh}\,,\\
4\pi P_{r}&=\frac{z(r-2m)+m/r-1}{r^{2}}\label{prh}\,,\\
4\pi P_{\perp}&=\left(1-\frac{2m}{r}\right)\mathcal{F}(z)+z\left(\frac{m}{r^{2}}-\frac{m'}{r}\right)\label{pperh},
\end{align}
where
\begin{eqnarray}
\mathcal{F}(z)=z'+z^{2}-\frac{z}{r}+\frac{1}{r^{2}}\,,
\end{eqnarray}
and the mass function $m(r)$ is defined by
\begin{eqnarray}\label{mF}
e^{-\lambda}=1-\frac{2m}{r}.
\end{eqnarray}
Note that, in contrast with the isotropic case reported in Ref. \cite{Lake2003}, three physical variables 
(principal stresses unequal) 
came into play. Therefore, in order for this protocol to work, two generating functions are needed; namely, $z(r)$ 
and $y(r)$. Incidentally,
if a less general technique of generation is required, only one input function is necessary, as shown in \cite{Lake2009}, 
to convert isotropic Newtonian static fluid spheres into general relativistic anisotropic static fluid spheres.
A different strategy consists in providing one generating function and an additional {\it ansatz} such as the conformally
flat condition, certain energy density distribution or a non local equation of state \cite{Herrera2008}, for example.

In this work we shall develop a protocol to generate all static and spherically symmetric anisotropic solutions for general 
relativistic polytropes. We note that the constraint introduced by the polytropic equation of state leaves room to only one 
generating function. Given the interest in polytropic equations of state for the relativistic community, the protocol
here developed could be of interest in several contexts (for applications of polytropes in astrophysics and related fields, see,
for example, Ref. \cite{book}).

In the next section we shall present the general equations and the algorithm which permits to construct all the solutions
previously mentioned and then we shall obtain the generating function for a specific solution previously considered in the
literature. The conformally flat and class I conditions are considered in subsequent sections before we give final remarks and conclusions.

\section{The algorithm}
In this section we develop an algorithm to obtain all static and spherically symmetric anisotropic 
solutions for general relativistic polytropes. For this purpose we shall
consider a line element parametrized in Schwarzschild-like coordinates as Eq. (\ref{linelement})
%
%
%
%
and write Einstein's equations (\ref{eq1}), (\ref{eq11}) and (\ref{eq111}) in terms of the generating function $y(r)$, given by (\ref{def2}), and the polytropic equation of state (\ref{ep2}). That is
%
%
%
%
%
\begin{eqnarray}
-\frac{1}{r^2}+\frac{y}{r^2}+\frac{y'}{r}&=&-8 \pi  \rho\,, \label{rhoy}\\
-\frac{1}{r^2}+\frac{y}{r^2}+\frac{\nu' y}{r}&=&8 \pi  K \rho ^{1+\frac{1}{n}}\,,\label{rhonuy} \\
\frac{\left(r \nu '+2\right) \left(y'+\nu'y\right)}{4r} +\frac{2 r\nu'' y}{4 r}&=& 8 \pi  P_{\perp}\,.\label{perpy}
\end{eqnarray}
Using \eqref{rhoy} and \eqref{rhonuy} we obtain

\begin{eqnarray}\label{nuprima}
\nu '=\frac{K r^{-2/n} \left(1-y-r y'\right)^{1+\frac{1}{n}}}{(8 \pi )^{1/n} ry}-\frac{y-1}{r y}.
\end{eqnarray}
Integrating the above equation, the metric function $\nu(r)$ turns to be 
%
\begin{equation}\label{nugfunction}
\nu = \int \frac{K \left(1-y-ry'\right)^{1+\frac{1}{n}}-(8 \pi r^{2})^{\frac{1}{n}}  (y-1)}{(8 \pi r^{n+2})^{\frac{1}{n}}y}dr\,.
\end{equation}
%
Then, replacing Eqs. \eqref{def2} and \eqref{nugfunction} in Eq. \eqref{linelement},
the line element takes the form
\begin{widetext}
\begin{equation}
ds^{2}=\exp\!\left[\int \frac{K \left(1-y-ry'\right)^{1+\frac{1}{n}}-(8 \pi r^{2})^{\frac{1}{n}}  (y-1)}{(8 \pi r^{n+2})^{\frac{1}{n}}y}dr\right]dt^{2}-\frac{1}{y}dr^{2}-r^{2}d\Omega^{2}\,.
\end{equation}
\end{widetext}

Finally, after a long but straightforward calculation, the tangential pressure (\ref{perpy}) can be expressed in terms of the\linebreak
generating function as
\begin{widetext}
\begin{align}
\label{pperp}
8\pi P_{\perp} \;=\;\, & \frac{1}{ \left(2r^{\frac{2+n}{n}}\right)^2\! y} 
\Bigg\{ \frac{1}{ (8\pi)^{2/n}\, n} 
\bigg[K^2 n \big(1-y-r y'\big)^{2\left(1+1/n\right)}
+ K (8 \pi r^2 )^{1/n} \big(1-y-r y'\big)^{1/n}\; \times\nonumber\\
& \hspace{.8cm} \times \Big\{2 n + 2 (2+n) y^2 + n r y'\big(r y'-3\big)
- y \Big[2 (1+n) \left(r^2 y''+2\right) + n r y'\Big] \Big\} \;\;+ \nonumber\\
& \hspace{.8cm} +\; (8\pi r^2)^{2/n}\, n \Big[r (y-1)y' + (y-2)y \Big] \bigg] + r^{4/n}\Bigg\} \,.
\end{align}
\end{widetext}

Therefore, the system is completely determined once the generating function, $y(r)$, is provided. 

It is worth noticing that the mass function, $m(r)$, is related to the generating function by means of
\begin{equation}\label{ym}
y \,=\, 1-\frac{2 m}{r}\,.
\end{equation}
Therefore, using the fact that Einstein's equations imply $4 \pi \rho = \frac{m'}{r^2}$, we have
\begin{equation}
\label{eqrho}
\rho \,=\, \frac{1-y-r y'}{8 \pi r^2}.
\end{equation}
Then, Eqs. (\ref{ep2}), (\ref{pperp}) and (\ref{eqrho}) express the physical variables in terms of the generating
function. Even more, using the polytropic equation of state (\ref{ep2}), and Eqs. (\ref{def1})
and (\ref{ym}), it can be seen that the anisotropic system described by (\ref{rhoh}), (\ref{prh}) and (\ref{pperh})
is reduced to the polytropic case under study. In this sense, the protocol for the polytropic fluid developed here
is consistent with the most general case reported in \cite{Herrera2008}. 


\section{Particular cases}
In this section we obtain the generating function for 
the polytropic models reported in 
\cite{Herrera2013a} and \cite{EFanisotropy}. It is worth noticing that although in both works the general relativistic polytrope problem is solved,
the main difference between them is the particular type of anisotropy employed.

\subsection{Model 1}

A particular solution reported in Ref. \cite{Herrera2013a} is based on the 
generalized Tolman-Oppenheimer-Volkoff equation in the form
\begin{equation}
\label{TOVA}
P'_{r}\,=\, -\frac{\nu'}{2}(\rho + P_{r})+\frac{2}{r}\Delta,
\end{equation}
together with a specific form for the anisotropy factor $\Delta=P_{\perp}-P_{r}=C f(P_{r},r)(\rho+P_{r})r^{N}$. 
In this previous equation, $C$ is a parameter
encoding the anisotropy, and $f$ and $N$ are certain function and number specific for each considered model
(see \cite{Cosenza1982} for details). In particular, if $f(P_{r},r)r^{N-1}=\frac{\nu'}{2}$ is assumed, then Eq. \eqref{TOVA} 
reads
\begin{equation}
\label{TOVAbis}
\frac{d P_{r}}{dr}=-h(\rho+P_{r})\frac{\nu'}{2},
\end{equation}
where $h=1-2C$. As noted in \cite{Herrera2013a}, after defining the usual variable entering the Lane-Emden equation, $w^{n}$, as
\begin{eqnarray}\label{rhorhoc}
\rho(r)=\rho_{c}w^{n}(r)
\end{eqnarray} 
where $\rho_{c}$ is the energy density at the center of the object, Eq. \eqref{TOVAbis} can be integrated, producing
\begin{eqnarray}
\label{eqint}
e^{\nu}=\left(1-\frac{2M}{r_{\Sigma}}\right)(1+ q_c w)^{-\frac{2(1+n)}{h}},
\end{eqnarray}
where $M$ and $r_{\Sigma}$ are the mass and radius of the object and $q_c=P_{c}/\rho_{c}$ with $P_{c}$ the radial 
pressure at its center.

Now, in order to obtain the generating function for this case, let us replace Eq. \eqref{rhorhoc}
in Eq. \eqref{rhoy} and solve for $y'(r)$ to obtain
\begin{eqnarray}\label{yprima}
y'=\frac{1 - y - 8\pi \rho_{c} r^{2} w^{n}}{r}\,. 
\end{eqnarray}

Combining \eqref{nuprima} and \eqref{yprima} we arrive to
\begin{eqnarray}\label{nug}
\nu'=\frac{1 - y + 8 \pi K r^2 \rho_{c}^{1+\frac{1}{n}} w^{1+n}}{r y}\,.
\end{eqnarray}
\\
Then, deriving Eq. \eqref{eqint} we obtain
\begin{eqnarray}\label{nuh}
\nu'= -\frac{2 q_c  K (1+n) w'}{h(1 + q_c w)}\,,
\end{eqnarray}
and finally, using \eqref{rhonuy} and \eqref{nuh}, the generating function takes the form
\begin{eqnarray}\label{y1}
y \,=\, \frac{h (1 + q_c w)\! \left(1 + 8\pi Kr^2 \rho_{c}^{1+\frac{1}{n}} w^{1+n} \right)}
{h + q_c  h w - 2q_c K (1+n) r w'}\,,
\end{eqnarray}
which completes the protocol. At this point, it is worth mentioning that although once the generating function is obtained 
the Einstein's equations are formally solved, as previously mentioned, the constraint given by Eq. \eqref{yprima}
must be fulfilled in order to obtain the metric function $\nu(r)$. 
We must emphasize that this procedure leads to the generalized Lane-Emden equation obtained in \cite{Herrera2013a} once the heuristic procedure applied to the anisotropy was introduced. This assures us that our function $y$, given previously, correctly generates the required solution.

\subsection{Model 2}
In this section, as a second example, we will concentrate in the double relativistic polytrope model introduced in \cite{EFanisotropy}. In this case, the equation of state for each pressure reads
 \begin{eqnarray}
P_{r} &=& K_{r}\rho^{1+\frac{1}{n_{r}}}\,,\\ 
P_{\perp} &=& K_{\perp}\rho^{1+\frac{1}{n_{\perp}}}\,,
\end{eqnarray}
with $\rho=\rho_{c}w^{n_{r}}$. The anisotropy factor can be written as
\begin{eqnarray}
\Delta=\rho_{c}q_{c}\omega^{n_{r}}(
\omega^{\theta}-\omega)\,,
\end{eqnarray}
where $\theta=n_{r}/n_{\perp}$ and $q_{c}=P_{rc}/\rho_{c}$. It can be shown that (see \cite{EFanisotropy} for details) 
\begin{eqnarray}\label{eso}
e^{\nu}=\frac{1-\frac{2M}{r_{\Sigma}}}{(1+q_{c}w)^{2(1+n_{r})}e^{4q_{c}G(r)}}\,,
\end{eqnarray}
with
\begin{eqnarray}
G(r) \,= \int\limits_{r}^{r_{\Sigma}}
\frac{w^{\theta}-w}{1+q_{c}w}\,dr' \,.
\end{eqnarray}

Deriving expression (\ref{eso}) we obtain
\begin{eqnarray}
\nu'\;=\; 2 q_{c}\frac{2 w^{\theta} - 2w - (1+n_{r}) w'}
{1 + q_c w}\;.
\end{eqnarray}
Finally, replacing in (\ref{rhonuy}) the generating function reads
\begin{eqnarray}\label{y2}
y \;=\; \frac{(1+q_{c} w) \left(1 + 8\pi P_{rc} r^2 w^{1+n_{r}}\right)}
{1 + q_{c}(1 - 4 r)w + 4q_{c}r w^{\theta} - 2(1 + n_{r})q_{c} r w'  }\,.\;\;
\end{eqnarray}

Before concluding this section, we would like to emphasize that Eqs. (\ref{y1}) and (\ref{y2}) correspond to formal expressions for the generating function, $y$, in the sense that they lead to the corresponding Lane-Emden equations 
which must be solved. Even more, given the high non--linearity of the polytropic equations, obtaining an analytical solution  for $y$ could be a difficult task. Therefore, in the following sections we propose some geometrical constraints which close the system and allow to obtain the generating function numerically.


\section{Conformally flat condition}

 A conformally flat model with polytropic equation of state was discussed in \cite{herrera2014}, so it is to be expected that this case is portrayed within our formalism. Conformally flat anisotropic spheres have vanishing Weyl tensor. In the spherically symmetric case, it can be shown that all non-vanishing components of it, $C_{\alpha\beta\gamma\delta}$, can be expressed through the component $C^3_{\;\,232}$. So, the Weyl tensor is described by a single function that must be canceled leading to a linear differential equation for $\nu'$, which can be integrated \cite{Herrera2001}. In this section we provide the conformally flat 
condition in order to close the system and find a solution for the generating function $y$. 

As it is well known, in terms of the standard parametrization (\ref{linelement}) the conformally flat condition reads
\begin{eqnarray}\label{scf}
\frac{e^{\lambda} }{r^2}-\frac{1}{r^2}+\frac{1}{4} \lambda ' \nu '-\frac{\lambda '-\nu '}{2 r}-\frac{\nu ''}{2}-\frac{1}{4} \nu '^2=0.
\end{eqnarray}
Now, after introducing the generating function by
 $e^{-\lambda}=y$, Eq. (\ref{scf}) can be written as
\begin{equation}\label{cfy}
r \left(r \nu '-2\right) y'+y \left(r \left(2 r \nu ''+\nu ' \left(r \nu '-2\right)\right)+4\right)-4=0.
\end{equation}
Next, using (\ref{nuprima}) and its derivative, Eq. (\ref{cfy}) can be expressed in terms of only the generating function and it can be numerically solved after a suitable choice of the boundary conditions and parameters involved are made.

\begin{figure}[h!]
	\centering
	\includegraphics[scale=0.49]{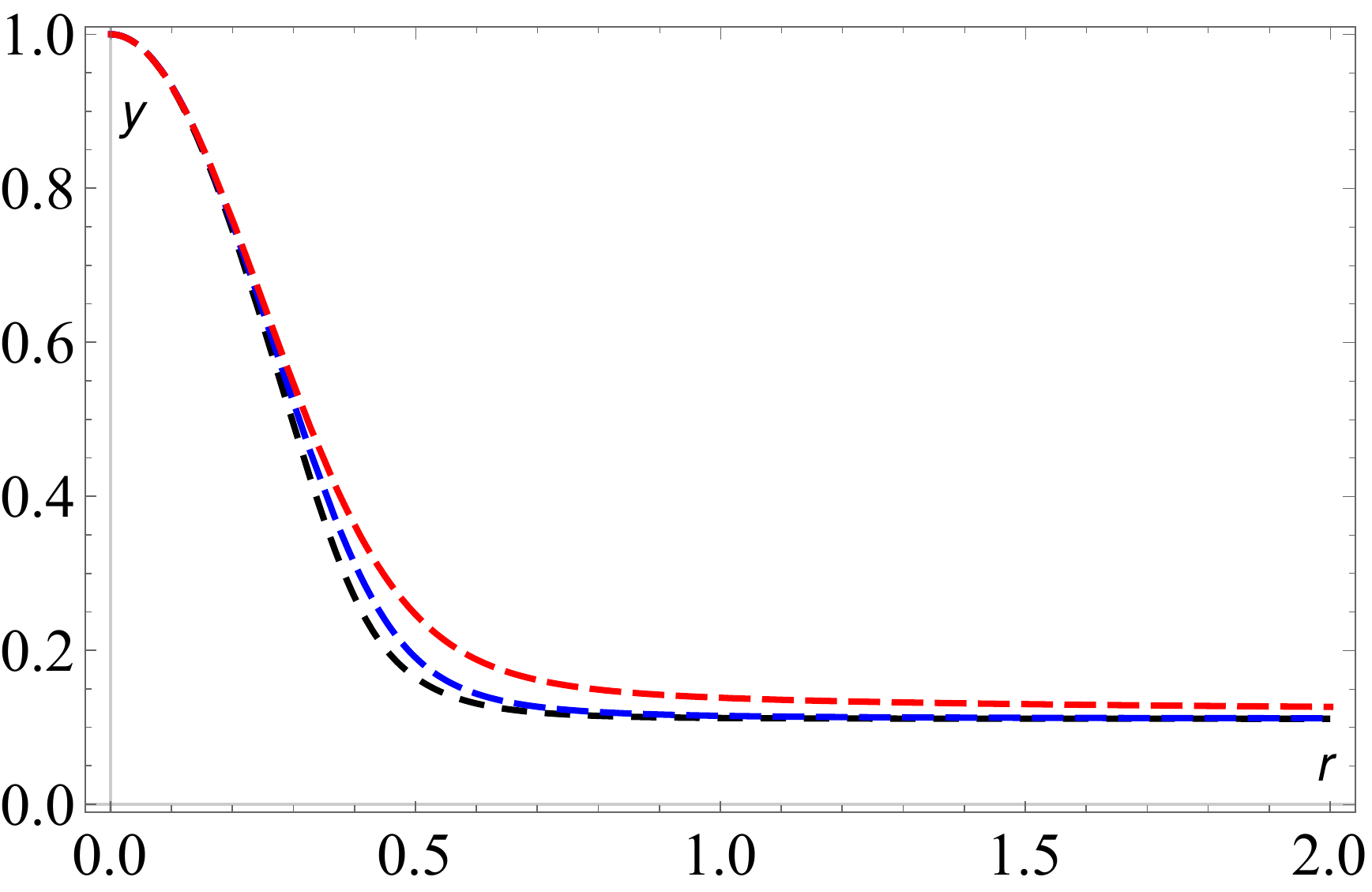}
	\caption{Generating function, $y$, as a function of $r$ for $K=0.5$ and 
		$n=0.5$ (black line), $n=1$ (blue line) and 
		$n=3$ (red line)}
	\label{y}
\end{figure}

In figure \ref{y} we show the behaviour of $y$ for the values of the parameters indicated in the caption. Note that the generating function is a monotonously decreasing function, as expected.


\section{Class I condition}

\begin{figure}[h!]
	\centering
	\includegraphics[scale=0.49]{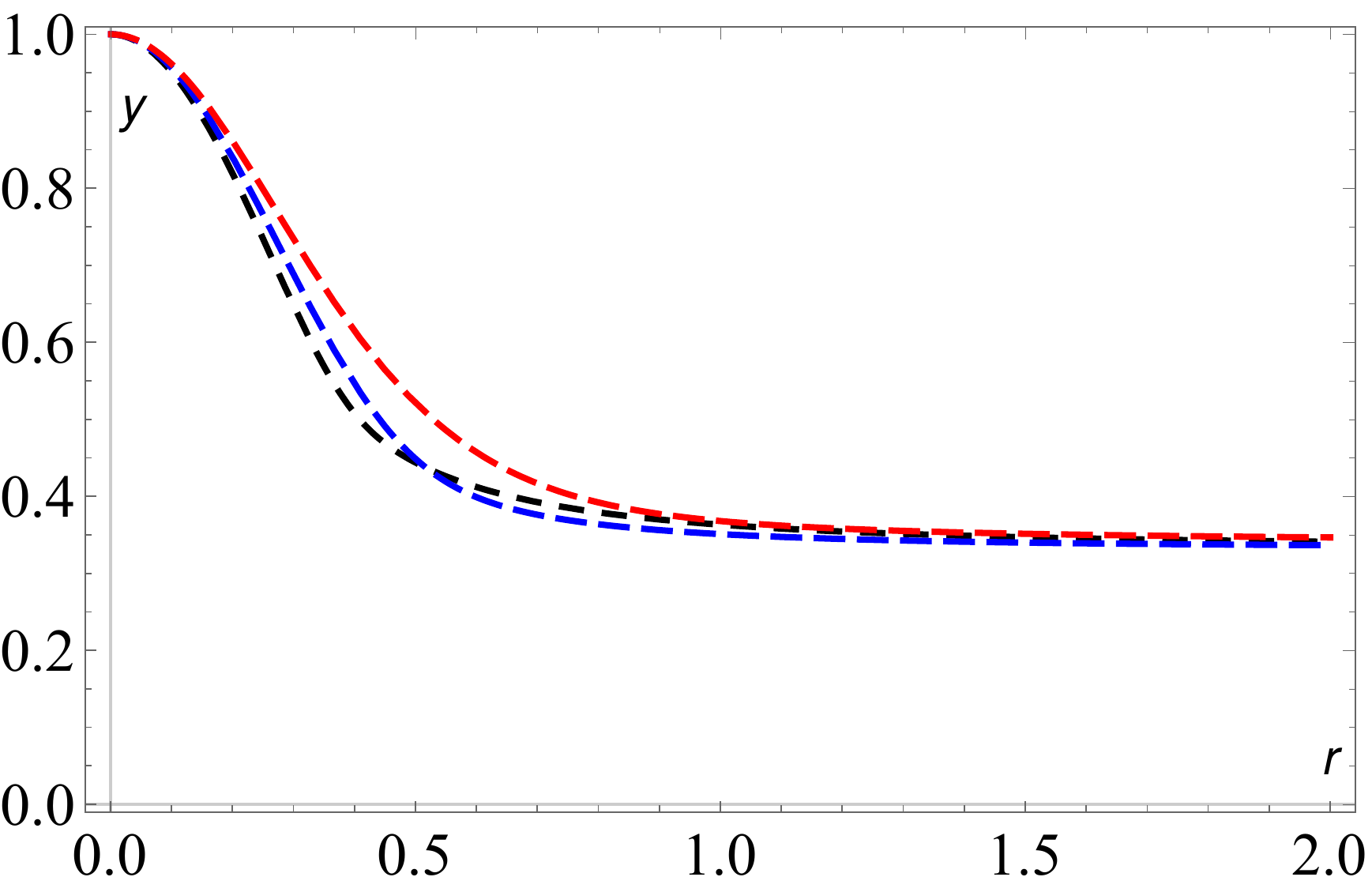}
	\caption{Generating function, $y$, as a function of $r$ for $K=0.5$ and 
		$n=0.5$ (black line), $n=1$ (blue line) and 
		$n=3$ (red line)}
	\label{y_class1}
\end{figure}

Embedding of four-dimensional space-times into higher dimensions is an invaluable tool in generating both cosmological and astrophysical models \cite{MM}. The necessary and sufficient condition required for a spherically symmetric space-time to be of class one to generate exact solutions of the interior Einstein's field equations is the Karmarkar condition \cite{karmarkar}. This reduces the gravitational behavior of the model to a single metric function. As it is well known, for spherically symmetric space--times
the class I condition reads ($R_{\theta\phi\theta\phi}\neq0$) \cite{eisland,karmarkar}
\begin{equation}\label{eq15}
R_{t\theta t\theta}R_{r\phi r\phi}=R_{trtr}R_{\theta\phi\theta\phi}+R_{r\theta t \theta}R_{r\phi t\phi},    
\end{equation}
which leads to
\begin{equation}\label{eq16}
2\frac{\nu^{\prime\prime}}{\nu^{\prime}}+\nu^{\prime}=\frac{\lambda^{\prime}e^{\lambda}}{e^{\lambda}-1},    
\end{equation}
with $e^{\lambda}\neq 1$. Then, using $e^{-\lambda}=y$ and (\ref{nuprima}) together with its derivative we arrive to a differential equation for $y$ that can be integrated numerically for several values of $n$ and some selected values for $K$. The results are shown in figure \ref{y_class1}, where it can be seen that the generating function $y$ is a monotonously decreasing function for the selected parameters.

\section{Final Remarks and conclusions}

In this work we have developed an algorithm to construct all static spherically symmetric anisotropic solutions for general relativistic polytropes in terms of one generating function. 
The protocol here developed could be useful for dealing with several phenomena in which anisotropic polytropes appear, with special emphasis in the physics of compact objects. It is important to note that our findings can be considered as particular cases of \cite{Herrera2008}. 

To illustrate the algorithm, we have followed two routes. 
In the first case, we have provided two solutions to the field equations in order for the generating function to be formally obtained. The first one is related with the general and detailed formalism to model polytropic general relativistic stars with anisotropic pressure \cite{Herrera2013a}. In this case, a heuristic model based on an ansatz to obtain anisotropic matter solutions from known solutions for isotropic matter was adopted. Within our scheme, we have been able to connect in Model $1$ with that result, which allows us to arrive to the Lane-Emden equation presented in \cite{Herrera2013a}. In the same way, in Model $2$ we have related our scheme with a recent work describing a general framework for modeling anisotropic relativistic polytropes, when both pressures ($P_r$ and $P_{\perp}$) satisfy a polytropic equation of state (the general relativistic double polytrope for anisotropic matter) \cite{EFanisotropy}. Both examples demonstrate that those cases are, indeed, contained within our general development. 

In the second case, we have provided extra constraints to close the system. In particular, we have chosen to work with geometrical constraints such as the conformally flat and Karmarkar conditions. We have found that, although in both cases a numerical analysis is required given the high non--linearity of the equations, the obtained function is monotonously decreasing as a function of the radial coordinate, as expected. In this regard, it could be interesting to explore other geometrical constraints which could allow one to obtain analytical solutions. This an other aspects are currently under study.

\section{Acknowledgments}
P. B. is funded by the Beatriz Galindo contract BEAGAL 18/00207 (Spain).

\end{document}